\documentclass[journal]{IEEEtran}
 \usepackage{natbib}
\usepackage{theorem,algorithm,algorithmic}
\usepackage{amssymb,amsmath}
\usepackage{setspace,ulem}

\usepackage{psfig,epsfig,amssymb,alltt}
\usepackage{latexsym}
\usepackage{graphicx}
\graphicspath{{PS/}}
\usepackage{psfig,amssymb,alltt}

\ifCLASSINFOpdf
\else
\fi
 
\newcommand{\be}{\begin{eqnarray}}
\newcommand{\ee}{\end{eqnarray}}

\newcommand{\bitem}{\begin{itemize}}
\newcommand{\eitem}{\end{itemize}}

\begin{document}
%
\title{Astronomical Data Analysis and Sparsity: from Wavelets to Compressed Sensing}

\author{Jean-Luc Starck and Jerome Bobin, 
\thanks{J.-L. Starck is with the Laboratoire AIM (UMR 7158), CEA/DSM-CNRS-Universit\'e Paris Diderot,   IRFU, SEDI-SAP, Service d'Astrophysique,  Centre de Saclay,  F-91191 Gif-Sur-Yvette cedex, France.}
\thanks{J. Bobin is  with the department of Applied and Computational Mathematics (ACM), California Institute of Technology, M/C 217-50, 1200 E.California, PASADENA CA-91125, USA.}
}

\markboth{Astronomical Data Analysis}%
{Shell \MakeLowercase{\textit{et al.}}: Bare Demo of IEEEtran.cls for Journals}


\maketitle

\begin{abstract}
Wavelets have been used extensively for several years now in astronomy for
many purposes, ranging from data filtering and deconvolution, to star and
galaxy detection or cosmic ray removal. More recent sparse representations
such ridgelets or curvelets have also been proposed for the detection of anisotropic features
such cosmic strings in the cosmic microwave background.
We review in this paper a range of methods based on sparsity that have been proposed for
astronomical data analysis. We also discuss what is  the impact of Compressed Sensing, the new sampling theory,
in astronomy for collecting the data, transferring them to the earth  or reconstructing an image from
incomplete measurements.
\end{abstract}


%
\IEEEpeerreviewmaketitle

\section{Introduction}
 The wavelet transform (WT) has  been extensively used in astronomical data 
analysis during the last ten years. A quick search with ADS 
(NASA Astrophysics Data System, adswww.harvard.edu)
\index{ADS}
shows that around 1000 papers contain the keyword ``wavelet'' in their abstract, and this 
holds for all astrophysical domains, from study of the sun through to 
CMB (Cosmic Microwave Background) analysis \cite{starck:book06}.
 This broad success of the wavelet transform is due to the fact 
that astronomical data generally gives rise to 
complex hierarchical structures, often described as fractals. 
Using multiscale approaches such as the wavelet transform, 
an image can be decomposed into components at different scales, and the
wavelet transform is therefore well-adapted to the study of astronomical data. 
Furthermore, since noise in the physical sciences is often not Gaussian, 
modeling in wavelet space of many kind of noise -- Poisson noise, 
combination of Gaussian and Poisson noise components, non-stationary 
noise, and so on --  has been a key motivation for the use of wavelets in astrophysics.

If wavelets represent well isotropic features, they are far from optimal for analyzing anisotropic objects.
This has motived other constructions such the curvelet transform \cite{cur:candes99_1}. More generally, the best data decomposition
 is the one which leads to the sparsest  representation,  i.e.  few coefficients have a large magnitude, while most of them are close to zero.
Hence, for specific astronomical data set containing edges (planetary images, cosmic strings, etc), curvelets should be preferred  to wavelets.

In this paper, we review a range of astronomical data analysis methods based on sparse representations.
We first introduce the Undecimated Isotropic Wavelet Transform (UIWT) which is the most popular WT algorithm 
in astronomy.  We show how the signal of interest can be detected in wavelet space using a noise modeling, allowing us to build
the so-called {\it multiresolution support}. Then we present in~\ref{sect_appli}  how this multiresolution support can used for restoration 
applications. 
In section~\ref{sect_curv}, another representation, the curvelet transform, is introduced, which is well adapted to
 anisotropic structure analysis. Combined together, the wavelet and the curvelet transforms are very powerful 
 to detect and discriminate very faint features. We give an example of application for cosmic string detection.
 Section~\ref{sect_cs} describes the compressed sensing theory which is strongly related to sparsity, and presents its impacts in astronomy,
 especially for spatial data compression.

 \section{The Isotropic Undecimated Wavelet Transform}
\label{sec:iuwt}

\begin{figure}
\centerline{
\hbox{
}}
\caption{Galaxy NGC 2997.}
\label{fig_galaxy}
\index{data!NGC 2997}
\end{figure}

\begin{figure*}[htb]
\centerline{
\hbox{
}}
\caption{Wavelet transform of NGC 2997 by the IUWT. The co-addition of these six images reproduces exactly the original image.}
\label{pl_pave_gala}
\index{data!NGC 2997}
\end{figure*}

The Isotropic  undecimated wavelet transform    (IUWT) \cite{Starck2006}  
decomposes  an $n \times n$ image $c_0$ into a coefficient set $W = \{w_1, \dots, w_J, c_J\}$,  
as a superposition of the form
\[
c_0[k,l] =  c_{J}[k,l] + \sum_{j=1}^{J} w_j[k,l],
\]
where $c_{J}$ is a coarse or smooth version of the original image $c_0$
and $w_j$ represents `the details of $c_0$' at scale $2^{-j}$ (see
Starck et al.\cite*{starck:book98,starck:book02} for more information). Thus, the
algorithm outputs $J+1$ sub-band arrays of size $n \times n$. (The
present indexing is such that $j = 1$ corresponds to the finest scale
(high frequencies)).

Hence, we have a {\it multi-scale pixel representation}, i.e. each
pixel of the input image is associated with a set of pixels of the
multi-scale transform. This wavelet transform is very well adapted to
the detection of isotropic features, and this explains  
its success for astronomical image processing, where the data contain
mostly isotropic or quasi-isotropic objects, such as stars, galaxies or
galaxy clusters.

The decomposition is achieved using the  filter bank $(h_{2\text{D}},g_{2\text{D}}=\delta-h_{2\text{D}},\tilde{h}_{2\text{D}}=\delta,\tilde{g}_{2\text{D}}=\delta)$ where $h_{2\text{D}}$ is the tensor product of  two 1D filters $h_{\mathrm{1D}}$.   The passage from one resolution to the next one is obtained using the  
``\`a trous'' algorithm~\cite{starck:book98}
\begin{eqnarray}
\label{eq:uwtdecomp}
c_{j+1}[k,l] &=&   \sum_m  \sum_n  h_{1D} [m]   h_{1D} [n]   c_{j}[k+2^{j}m, l+2^{j}n], \quad  \nonumber \\
w_{j+1}[k,l] &=& c_j[k,l]  - c_{j+1}[k,l] ~,
\end{eqnarray}
where $h_{1D}$ is typically a symmetric low-pass filter such as the $B_3$-Spline filter. 
  
 Fig.\ \ref{pl_pave_gala} shows IUWT of the galaxy
NGC 2997 displayed in Fig.~\ref{fig_galaxy} .  
Five wavelet scales are shown and the final smoothed plane (lower right).
The original image is given exactly by the sum of these
six images.  

\subsection{Example: Dynamic  range compression using the IUWt}

\begin{figure}[htb]
\vbox{ 
\centerline{ 
\hbox{
}} 
\centerline{ \hbox{
}} 
} 
\caption{Top left -- Hale-Bopp Comet image. Bottom  left -- 
histogram equalization results.
Bottom  right -- wavelet-log representations.}
\label{fig_halebopp_wavelet}
\end{figure}

Since some features in an image may be hard to detect by the human
eye due to low contrast, we often process the image before
visualization. Histogram equalization is certainly one the most
well-known methods for contrast enhancement. Images with a high
dynamic range are also difficult to analyze. For example,
astronomers generally visualize their images using a logarithmic look-up-table conversion.

Wavelets can be used to compress the dynamic range at all
scales, and therefore allow us to clearly see some very faint
features. For instance, the wavelet-log representation consists
of replacing $w_{j}[k,l]$ by $ \mathrm{sgn}(w_{j}[k,l]) \log(| w_{j}[k,l]|)$, leading to the
alternative image
\begin{eqnarray}
I_{k,l} = \log(c_{J,k,l}) +  \sum_{j=1}^{J} \mathrm{sgn}(w_{j}[k,l]) \log(\mid w_{j}[k,l]\mid + \epsilon)
\end{eqnarray}
where $\epsilon$ is a small number (for example $\epsilon=10^{-3}$).
Fig.\ \ref{fig_halebopp_wavelet} shows a Hale-Bopp Comet image (logarithmic representation)
(top left), its
histogram equalization (middle row), and its wavelet-log
representation (bottom). Jets clearly appear in the last
representation of the Hale-Bopp Comet image.

\subsection{Signal detection in the wavelet space}
\label{sect_noise}
   Observed data $Y$ in the physical sciences 
are generally corrupted by noise, which is often additive and which 
follows in many cases a Gaussian distribution, a Poisson distribution, or
a combination of both. It is important
to detect the wavelet coefficients which are ``significant'', i.e.   
the wavelet coefficients which have an absolute value too large to be 
due to noise. We defined the multiresolution $M$ of an image $Y$   by:
\begin{eqnarray}
M_{j}[k,l] = \left\{
  \begin{array}{ll}
  \mbox{ 1 } & \mbox{ if }   w_{j}[k,l]  \mbox{ is significant} \\
  \mbox{ 0 } & \mbox{ if }  w_{j}[k,l] \mbox{ is not significant}
  \end{array}
  \right.
\end{eqnarray}
where  $w_{j}[k,l]$ is the wavelet
coefficient of $Y$ at scale $j$ and at position $(k,l)$.
We need now to determine when a wavelet coefficient is significant.
For Gaussian noise, it is easy to derive an estimation
of the noise standard deviation $\sigma_j$ at scale $j$ from the 
noise standard deviation, which can be evaluated with good accuracy in
an automated way \cite{starck:sta98_3}.
To detect the significant wavelet coefficients, 
it suffices to compare the wavelet coefficients $w_{j}[k,l]$ to a threshold level
$t_j$. $t_j$ is generally taken equal to $K \sigma_j$, and 
 $K $ is chosen between 3 and 5. The value of 3 corresponds to a probability
 of false detection of $0.27\%$.
If $w_{j}[k,l]$ is small, then 
it is not significant and could be due to noise.  If
$w_{j}[k,l]$ is large, it is significant:
\begin{eqnarray}
\begin{array}{l}
\mbox{ if }  \mid w_{j}[k,l] \mid \ \geq \ t_j \ \ \mbox{ then } w_{j}[k,l] \mbox{ is 
significant } \\ 
\mbox{ if }  \mid w_{j}[k,l]  \mid \ < t_j \ \ \mbox{ then } w_{j}[k,l]  \mbox{ is not 
significant }
\end{array}
\end{eqnarray}


When the noise is not Gaussian, other strategies may be used:
\begin{itemize}
\item{\bf Poisson noise:}
if the noise in the data $Y$ is Poisson, the transformation 
\cite{rest:anscombe48} ${\cal A}(Y) = 2\sqrt{I + \frac{3}{8}}$
acts as if the data arose from a
Gaussian white noise model, with $\sigma = 1$, under the
assumption that the mean value of $I$ is sufficiently large.
However, this transform has some limits and it has been shown that 
it cannot be applied for data with less than 20 photons per pixel.
So for X-ray or gamma ray data, other solutions have to be chosen, which
manage the case of a reduced number of events or photons under 
assumptions of Poisson 
statistics

\item{\bf Gaussian + Poisson noise:}
the  generalization of  variance stabilization \cite{starck:mur95_2} is:
\begin{eqnarray*}
{\cal G}((Y[k,l]) = \frac{2}{\alpha} \sqrt{\alpha Y[k,l] + \frac{3}{8} \alpha^2 + \sigma^2 -
\alpha g}
 \end{eqnarray*}
where $\alpha$ is the gain of the detector, and $g$ and $\sigma$ are the mean and
the standard deviation of the read-out noise.  

\item{\bf Poisson noise with few events using the MS-VST}
For images with very few photons, one solution consists in using the Multi-Scale Variance Stabilization Transform (MSVST) \cite{starck:zhang07}.
The MSVST combines both the Anscombe transform and the IUWT in order to produce {\it stabilized} wavelet coefficients, i.e. coefficients corrupted by
a Gaussian noise with a standard deviation equal to 1. In this framework, wavelet cofficients are now calculated by:
{ 
\be
\label{eq:coupled:msvst}
\begin{tabular}{c}
\mbox{IUWT} \\
\mbox{+} \\
\mbox{MS-VST} \\
\end{tabular}
\left\{\begin{array}{lll}
c_j &=&     \sum_m  \sum_n  h_{1D} [m]   h_{1D} [n]   \\
      &  &  c_{j-1}[k+2^{j-1}m, l+2^{j-1}n]       \\
w_j &=& {\cal{A}}_{j-1}(c_{j-1}) -  {\cal{A}}_j(c_j)
\end{array}\right.
\ee
}
where ${\cal{A}}_j$ is the VST operator at scale $j$ defined by:
\begin{equation}
\label{eq:vstj:separated}
{\cal{A}}_j(c_j) = b^{(j)}  \sqrt{|c_j + e^{(j)}|}
\end{equation}
where  the variance stabilization constants $b^{(j)}$ and $e^{(j)}$ only depends on the filter $h_{1D} $ and the scale level $j$. They   
 can all be pre-computed once for any given $h$ \cite{starck:zhang07}. The multiresolution support is computed from the MSVST coefficients,
 considering a Gaussian noise with a standard deviation equal to 1.
 This stabilization procedure is also  invertible as we have: 
\begin{equation}
\label{eq:iuwt:inverse}
c_0 = {\cal{A}}_0^{-1}\left[{\cal{A}}_J(a_J) + \sum_{j=1}^J c_j\right]
\end{equation} 
\end{itemize}
For other kind of noise (correlated noise, non stationary noise, etc), other solutions  have been proposed to derive the multiresolution support \cite{starck:book06}. 
In next section, we show how the multiresolution support can be used for denoising and deconvolution.

\section{Restoration using the Wavelet Transform}
\label{sect_appli}
\subsection{Denoising}
The most used filtering method is the hard thresholding, which
 consists of setting to 0 all wavelet coefficients of $Y$
 which have an absolute value lower than a threshold $t_j$ 
\begin{eqnarray}
\tilde w_{j}[k,l]= \left\{
  \begin{array}{ll}
   w_{j}[k,l] & \mbox{ if }   \mid w_{j}[k,l]  \mid >  t_j \\
  0 &  \mbox{ otherwise }
  \end{array}
  \right.
\end{eqnarray}
More generally,  for a given sparse representation (wavelet, curvelet, etc) with its associate fast transform $ {\mathcal T}_w$ and fast reconstruction $ {\mathcal R}_w $, 
we can derive a hard thresholding denoising solution $X$ from the data $Y$, by first estimating the  
multiresolution support  $M $ using a given noise model, and  then calculating:
\begin{equation}
X =     {\mathcal R}_w   M   {\mathcal T}_w Y.
\end{equation}
We transform the data, multiply  the coefficients by the support and reconstruct  the solution.

The solution can however be improved considering the following optimization problem $ \min_{X} \parallel M ({\mathcal T}_w Y - {\mathcal T}_w X ) \parallel_2^2$ where $M$
is the multiresolution support of $Y$. A solution  can be obtained using the Landweber iterative scheme \cite{starck:sta95_1,starck:book98}: 
\begin{eqnarray}
{X}^{n+1} = {X}^n + {\mathcal R}_w  M \left[  {\mathcal T}_w Y   - {\mathcal T}_w  {X}^n \right]
\end{eqnarray}
If the solution is known to be positive, the positivity constraint can be 
introduced using the following equation:
\begin{eqnarray}
\label{Eq:projgrad}
X^{n+1} = P_{+}
 \left( 
X^n + {\mathcal R}_w M \left[  {\mathcal T}_w Y  - {\mathcal T}_w  X^n \right] \right)
\end{eqnarray}
where $P_{+}$ is the projection on the cone of non-negative images.

This algorithm allows us to constraint the residual to have a zero value inside the  multiresolution support   \cite{starck:book98}.  For astronomical
image filtering,  iterating improves significantly the results, especially for the photometry (i.e. the integrated number of photons in a given object).

\subsection{Deconvolution}

\begin{figure*}[htb]
\vbox{
\centerline{  
\hbox{
}}}
\caption{Simulated Hubble Space Telescope  image of a distant cluster of 
galaxies.  Left: original, unaberrated and noise-free.
middle: input, aberrated, noise added.  Right,  wavelet restoration wavelet. }
\label{fig_caulet_freudling}
\end{figure*}

In a deconvolution problem,  $Y = H X + N$, when the sensor is linear, $H$ is the block Toeplitz matrix.
Similarly to the denoising problem, the solution can   be obtained   minimizing $ \min_{X} \parallel M {\mathcal T}_w (Y - H X)  \parallel_2^2$ under a positivity constraint, 
leading to the 
Landweber iterative scheme \cite{starck:sta95_1,starck:book98}: 
\begin{eqnarray}
\label{Eq:projdecgrad}
X^{n+1} = P_{+}
 \left( 
X^n + H^t  {\mathcal R}_w M   {\mathcal T}_w  \left[   Y  - H   X^n \right] \right)
\end{eqnarray}
Only coefficients that belong 
to the  multiresolution  support   are kept, while the others are set to zero \cite{starck:sta95_1}.  
At each iteration, the multiresolution support $M$ can be  updated by selecting new coefficients in the wavelet transform of the residual which have an absolute value larger than a given threshold. 
 
\subsection*{Example}
A simulated Hubble Space Telescope image of a distant cluster of galaxies is shown in 
Fig.~\ref{fig_caulet_freudling},   middle.
  The simulated data are shown
in Fig.~\ref{fig_caulet_freudling}, left.  Wavelet deconvolution solution is shown  Fig.~\ref{fig_caulet_freudling}, right.
The method is stable for any kind of point spread function, and any kind of noise modeling can be considered.


\subsection{Inpainting}
 
Missing data are a standard problem in astronomy. They can be due to bad pixels, or image  area   we consider as problematic due to calibration or observational problems.  These masked area lead to many difficulties for post-processing, 
especially to estimate statistical information such the power spectrum or the bispectrum. The inpainting technique consists in filling the gaps.
The classical image inpainting problem can be defined as follows. 
Let $X$ be the ideal complete image, $Y$ the observed incomplete image and $L$ the binary mask (i.e. $L[k,l] = 1$ if we 
have information at pixel $(k,l)$, $L[k,l] = 0$ otherwise). In short, we have: $Y = L X$. 
Inpainting consists in recovering $X$ knowing $Y$ and $L$.
  
Noting $|| z ||_0$ the $l_0$ pseudo-norm, i.e. the number of non-zero entries in $z$ and $|| z ||$ the classical $l_2$ norm (i.e. $ || z ||^2 = \sum_k (z_k)^2 $), we thus want to minimize:
\begin{equation}
\min_X  \| \Phi^T X \|_0 \quad   \text{subject to}  \quad   \parallel Y - L X  \parallel_{\ell_2} \le \sigma,
\label{minimisation}
\end{equation}
where $\sigma$ stands for the noise standard deviation in the noisy case. 
It has also been shown that if $X$ is sparse enough, the $l_0$ pseudo-norm can also be replaced by the convex $l_1$ norm (i.e. $ || z ||_1 = \sum_k | z_k | $) \citep{cur:donoho_01b}. The solution of such an optimization task can be obtained through an iterative thresholding algorithm called MCA \citep{inpainting:elad05,starck:jalal06} :
\begin{equation}
   X^{n+1} = \Delta_{\Phi,\lambda_n}(X^{n} + Y - L X^n)
\label{eqn_mca}
\end{equation}
where the nonlinear operator $\Delta_{\Phi,\lambda}(Z)$ consists in:
\begin{itemize}
\item decomposing the signal $Z$ on the dictionary $\Phi$ to derive the coefficients $\alpha = \Phi^T Z$.
\item threshold the coefficients: ${\tilde \alpha} = \rho(\alpha, \lambda)$, 
where the thresholding operator $\rho$   can either
be a hard thresholding (i.e. $\rho(\alpha_i, \lambda) = \alpha_i$ if $ | \alpha_i | > \lambda$ and $0$ otherwise)
 or a soft thresholding (i.e.
  $\rho(\alpha_i, \lambda) = \mathrm{sign}(\alpha_i) \mathrm{max}(0, | \alpha_i |  - \lambda)$). 
 The hard thresholding corresponds to the $l_0$ 
 optimization problem while the soft-threshold solves that for $l_1$.
\item reconstruct $\tilde Z$ from the thresholds coefficients ${\tilde \alpha}$.
\end{itemize}
The threshold parameter $\lambda_n$ decreases with the iteration number and it plays a role similar to the cooling parameter  
of the simulated annealing techniques, i.e. it allows the solution to escape from local minima. 
More details relative to this optimization problem can be found in \citep{CombettesWajs05,starck:jalal06}. For many dictionaries such as wavelets
or Fourier, fast operators exist to decompose the signal so that the iteration of eq.~\ref{eqn_mca} is very fast. It requires only to perform
at each iteration  a forward transform, a thresholding of the coefficients and an inverse transform. 

\subsubsection*{Example}
\begin{figure*}[htb]
\vbox{
\centerline{
\hbox{
}}
}
\caption{Left panel, simulated weak lensing mass map, middle panel, simulated mass map with the mask pattern of CFHTLS data, right panels, inpainted mass map. 
The region shown is $1^\circ$ x $1^\circ$.}
\label{inpainting}
\end{figure*}
The  experiment was conducted on a  simulated weak lensing mass map masked by a  typical mask patterns (see Fig. \ref{inpainting}). 
The left panel shows the simulated mass map and the middle panel show the masked map.  The result of the inpainting method  
is shown in the right panel. We note that the gaps are undistinguishable by eye. More interesting, it has been shown that, using the inpainted map, we 
can reach an accuracy of about  $1 \%$ for the power spectrum and $3 \%$ for the bispectrum  \cite{pires09}.

\section{From Wavelet to Curvelet}
 \label{sect_curv}
 
\begin{figure}
\centerline{
\hbox{
}}
\caption{A few first generation curvelets.}
\label{fig_ex_curve}
\end{figure}

The 2D curvelet  transform \cite{cur:candes99_1} was developed in an attempt to overcome some limitations inherent in former multiscale 
methods \emph{e.g.} the 2D wavelet, when handling smooth images with edges \textit{i.e.} singularities along smooth curves. 
 Basically, the curvelet dictionary is a multiscale 
pyramid of localized directional functions with anisotropic support obeying a specific parabolic scaling such that at scale $2^{-j}$, its length is $2^{-j/2}$ 
and its width is  $2^{-j}$. This is motivated by the parabolic scaling property of smooth curves. 
Other properties of the curvelet transform as well as decisive optimality results in approximation theory are reported in \citep{CandesDonohoCurvelets}. 
Notably, curvelets provide optimally sparse representations of manifolds which are smooth away from edge singularities along smooth curves.
Several digital curvelet transforms \citep{starck:sta01_3,cur:demanet06} have been proposed which attempt to preserve the essential properties 
of the continuous curvelet transform and several papers report on their successful application in astrophysical experiments \cite{starck:sta02_3,starck:sta03_1,starck:sta05_2}. 

Fig.~\ref{fig_ex_curve} shows a few curvelets at different scales,
orientations and locations.


\subsection*{Application to the detection of cosmic strings}

Some applications require the use of sophisticated statistical tools in 
order to detect a very faint signals, embedded in noise.
An interesting case is the detection of non-Gaussian signatures in Cosmic Microwave Background (CMB).
which is  of great interest for cosmologists.  Indeed, the non-Gaussian
signatures in the CMB can be related to very fundamental questions
such as the global topology of the universe \cite{riazuelo2002}, 
superstring theory, topological defects such as cosmic strings
\cite{bouchet88}, and multi-field inflation \cite{bernardeau2002}.  The
non-Gaussian signatures can, however, have a different but still
cosmological origin. They can be associated with the
Sunyaev-Zel'dovich (SZ) effect \cite{sunyaev80} (inverse Compton
effect) of the hot and ionized intra-cluster gas of galaxy clusters
\cite{gauss:aghanim99}, with the gravitational lensing by
large scale structures, or with the reionization
of the universe \cite{gauss:aghanim99}.  They may also be
simply due to foreground emission, or to   non-Gaussian instrumental noise and systematics .

All these sources of non-Gaussian signatures might have different
origins and thus different statistical and morphological
characteristics.  It is therefore not surprising that a large number
of studies have recently been devoted to the subject of the detection
of non-Gaussian signatures.  In \cite{gauss:aghanim03,starck:sta03_1}, it was shown that the wavelet
transform was a very powerful tool to detect the non-Gaussian
signatures. Indeed, the excess kurtosis (4th moment) of the wavelet
coefficients  outperformed all the other methods 
(when the signal is characterized by a non-zero 4th moment).
 
Finally, a major issue of the non-Gaussian studies in CMB remains
our ability to disentangle all the sources of non-Gaussianity from one
another. It has been shown it was
possible to separate the non-Gaussian signatures associated with
topological defects (cosmic strings) from those due to the Doppler effect
of moving clusters of galaxies (i.e. the kinetic Sunyaev-Zel'dovich effect),  
both dominated by a Gaussian CMB
field, by combining the excess kurtosis derived from both the wavelet
and the curvelet transforms \cite{starck:sta03_1}. 

The wavelet transform is suited to spherical-like
sources of non-Gaussianity, and a curvelet transform  is suited to 
structures representing 
 sharp and elongated structures such as cosmic strings.  The combination of these transforms
 highlights the presence of the cosmic strings in a mixture
CMB+SZ+CS. Such a combination   gives information about the
nature of the non-Gaussian signals.   
The sensitivity of each transform to a particular shape makes it a very
strong discriminating tool \cite{starck:sta03_1,starck:jin05}.

\begin{figure}[htb]
\centerline{
\vbox{
}}
\caption{Top, primary Cosmic Microwave Background anisotropies (left) and 
kinetic Sunyaev-Zel'dovich fluctuations (right).
Bottom, cosmic string simulated map (left) and simulated observation
containing the previous three components (right).
 The wavelet function is overplotted on the 
  Sunyaev-Zel'dovich map and the curvelet function is overplotted on 
the cosmic string map.}
\label{fig_cmb}
\end{figure}
In order to illustrate this, we show in
Fig.~\ref{fig_cmb} a set of simulated maps. Primary CMB, kinetic SZ
and cosmic string maps are shown respectively in Fig. \ref{fig_cmb}
top left, top right and bottom left. The ``simulated observed
map", containing the three previous components, is displayed in
Fig. \ref{fig_cmb} bottom right. The primary CMB anisotropies dominate all the
signals except at very high multipoles (very small angular scales).
The wavelet function is overplotted on the 
kinetic Sunyaev-Zel'dovich map and the curvelet function is overplotted on cosmic string map.

\section{Compressed Sensing}
\label{sect_cs}
\subsection{Compressed Sensing in a nutshell}
Compressed sensing (CS) \cite{CRT:cs,donoho:cs} is a new sampling/compression
theory based on the revelation that one can exploit sparsity or
compressibility when acquiring signals of general interest, and that
one can design nonadaptive sampling techniques that condense the
information in a compressible signal into a small amount of
data.  The gist of Compressed Sensing (CS) relies on two fundamental properties~:
\begin{enumerate}
\item{\textit{ Compressibility of the data~:} }  The signal $X$ is said to be \textit{compressible}  if it exists a dictionary $\Phi$ where the  coefficients $\alpha = \Phi^T X$,   obtained after decomposing $X$ on $\Phi$,  are sparsely distributed.
\item{\textit{ Acquiring incoherent measurements~: }} 
In the Compressed Sensing framework, the signal $X$ is not acquired directly; one then acquires a signal $X$ by collecting data of the form $Y = A X + \eta$~: $A$ is an $m \times n$ (with $m < n$) ``sampling'' or measurement matrix, and $\eta$ is a noise term. Assuming $X$ to be sparse, the incoherence of $A$ and $\Phi$ (\textit{e.g.} the Fourier basis and the Dirac basis) entails that the information carried by $X$ is diluted in all the measurements $Y$. Combining the incoherence of $A$ and $\Phi$ with the sparsity of $X$ in $\Phi$ makes the decoding problem tractable.
\end{enumerate}
In the following, we choose the measurement matrix $A$ to be a submatrix of an orthogonal matrix $\Theta$~: the resulting measurement matrix is denoted $\Theta_{\Lambda}$ and obtained by picking a set of columns of $\Theta$ indexed by $\Lambda$; $\Theta_{\Lambda}$ is obtained by subsampling the transformed signal $\Theta X$. In practice, when $\Theta$ admits a fast implicit transform (\textit{i.e.} discrete Fourier transform, Hadamard transform, noiselet transform), the compression step is very fast and made reliable for on-board satellite implementation.\\
A standard approach in CS attempts to reconstruct $X$ by solving
\begin{equation}\label{eq:cs}
\min_\alpha \| \alpha \|_{\ell_1} \mbox{ s.~t. } \|Y - \Theta_{\Lambda} \Phi \alpha \|_{\ell_2} < \epsilon
\end{equation}
where $\epsilon^2$ is an estimated upper bound on the noise power.

\subsection{Compressed sensing for the Herschel data}

The Herschel/PACS mission of the European Space Agency (ESA) is facing with a strenuous compression dilemma~: it needs a compression rate equal to $\rho = 1/N$ with $N=6$. A first approach has been proposed which consists in averaging $N=6$ consecutive images of a raster scan and transmitting the final average image. Nevertheless, doing so with high speed raster scanning leads to a dramatic loss in resolution.
In \cite{starck:bobin_herschel}, we emphasized on the redundancy of raster scan data~: $2$ consecutive images are almost the same images up to a small shift $\delta$. Then, jointly compressing/decompressing consecutive images of the same raster scan has been put forward to alleviate the Herchel/PACS compression dilemma. The problem then consists in recovering a single image $X$ from $N$ compressed and shifted noisy versions of $X$~:\\
\begin{equation}
\forall i \in \{1,\cdots,N\}; \quad X_i = \mathcal{T}_{\delta_i}\left(X\right) + \eta_i
\end{equation}
where $\mathcal{T}_{\delta_i}$ is an operator that shifts the original image $X$ with a shift $\delta_i$. The term $\eta_i$ models instrumental noise or model imperfections. According to the compressed sensing framework, each signal is projected onto the subspace ranged by $\Theta$. Each compressed observation is then obtained as follows~:
\begin{equation}
\forall i \in \{1,\cdots,N\}; \quad Y_i = {\Theta_i}_{\Lambda_i} X_i
\end{equation} 
where the sets $\{\Lambda_i\}$ are such that the union of all the measurement matrices $[\Theta_{\Lambda_1},\cdots,\Theta_{\Lambda_1}]$ span $\mathbb{R}^n$.
In practice, the subsets $\Lambda_i$ are disjoint and have a cardinality $m = \lfloor n/N \rfloor$. When there is no shift between consecutive images, these conditions guarantee that the signal $X$ can be reconstructed univocally from $\{Y_i\}_{i=1,\cdots,N}$, up to noise.
The decoding step amounts to seeking the signal $x$ as follows~:
\begin{equation}
\label{eq:l1_pacs}
\min_\alpha \| \alpha \|_{\ell_1} \mbox{ s.~t. } \sum_{i=1}^N \|Y_i - \Theta_{\Lambda_i} \Phi \alpha \|_{\ell_2} < \sqrt{N}\epsilon
\end{equation} 
The solution of this optimization problem can be found via an iterative thresholding algorithm (see \cite{starck:bobin_herschel})~:
\begin{equation}\label{eqn_cs}
   X^{n+1} = \Delta_{\Phi,\lambda_n} (X^{n} + \mu_{\Theta} \sum_{i=1}^N \Theta_{\Lambda_i}^T \left(Y_i - \Theta_{\Lambda_i} X^n) \right)
\end{equation}
where the nonlinear operator $\Delta_{\Phi,\lambda}(Z)$ is defined in Equation~\ref{eqn_mca} and the step-size $\mu_{\Theta_{\Lambda}} < 2/\sum_i\|\Theta_{\Lambda_i}^T \Theta_{\Lambda_i} \|_2$. Similarly to the MCA algorithm, the threshold $\lambda_n$ decreases with the iteration number towards the final value~: $\lambda_f$; a typical value is $\lambda_f = 2-3 \sigma$. This algorithm has been shown to be very efficient for solving the problem in Equation~\ref{eq:cs} in \cite{starck:bobin_herschel}.

 \begin{center}
\begin{figure}[htb]
$$
\begin{array}{cc}
\end{array}
$$
\vspace{-0.1in} 
\caption{
{Top left :} Original image. 
{Top right :} First input noisy map (out of 6). The PACS data already contains approximately Gaussian noise. 
{Bottom left :} Mean of the $6$ input images. 
{Bottom right :} Reconstruction from noiselet-based CS projections. The iterative algorithm has been used with $100$ iterations.} 
\label{fig:res_pacs}
\end{figure}
\end{center}

\paragraph{Illustration}
We compare two approaches to solve the Herschel/PACS compression problem~: i) transmitting the average of $6$ consecutive images (MO6), ii) compressing $6$ consecutive images of a raster scan and decompressing using Compressed Sensing. Real Herschel/PACS data are complex~: the original datum $X$ is contaminated with a slowly varying ``flat field" component $c_f$. In a short sequence of $6$ consecutive images, the flat field component is almost fixed. In this context, the data $\{x_i\}_{i=0,\cdots,1}$ can then be modeled as follows~:
\begin{equation}
X_i = \mathcal{T}_{\delta_i}\left(X\right) + \eta_i + c_f
\end{equation}
If $c_f$ is known (which will be the case in the forthcoming experiments), $\mathcal{T}_{\delta_i}\left(X^{(n)}\right)$ is replaced by $\mathcal{T}_{\delta_i}\left(X^{(n)}\right) + c_f$ in Equation~\ref{eqn_cs}. The data have been designed by adding realistic pointwise sources to real calibration measurements performed in mid-2007. In the following experiment, the sparsifying dictionary is $\Phi$ is an undecimated wavelet tight frame and the measurement matrices are submatrices of the noiselet basis \cite{Coifman01noiselets}.\\
The top-left picture of Figure~\ref{fig:res_pacs} features the original signal $X$. In the top-right panel of Figure~\ref{fig:res_pacs}. The ``flat field" component overwhelms the useful part of the data so that $x$ has at best a level that is $30$ times lower than the ``flat field" component. The MO6 solution (\textit{resp.} the CS-based solution) is shown on the left (\textit{resp.} right) and at the bottom of Figure~\ref{fig:res_pacs}. We showed in \cite{starck:bobin_herschel} that Compressed Sensing provides a resolution enhancement that can reach $30\%$ of the FWHM of the instrument's PSF for a wide range of signal intensities (\textit{i.e.} flux of $X$).\\
This experiment illustrates the reliability of the CS-based compression to deal with real-world data compression. The efficiency of Compressed Sensing applied to the Herschel/PACS data compression relies also on the redundancy of the data~: consecutive images of a raster scan are fairly shifted versions of a reference image. The good perfomances of CS is obtained by merging the information of consecutive images. The same \textbf{data fusion} scheme could be used to reconstruct with high accuracy wide sky areas from full raster scans.

\section{Conclusion}

By establishing a direct link between sampling and sparsity, compressed sensing had a huge
impact in many scientific fields, especially in astronomy. We have seen that CS could offer an elegant solution to the Herschel data transfer problem.  
By emphasing so rigorously the importance of sparsity, compressed sensing has also shed light on 
all work related to sparse data representation (such as the wavelet 
transform, curvelet transform, etc.). Indeed, a signal is generally not  sparse in direct space (i.e.\ pixel space), 
but it can be very sparse after being decomposed on a specific set of 
functions. For inverse problems,   compressed sensing gives 
a strong theoretical support for  methods which seek a sparse solution, since such a solution may be 
(under appropriate conditions) the exact one. Similar results are hardly accessible with other regularization methods.
This explain why wavelets and curvelets are so successful for astronomical image denoising, deconvolution and inpainting.

\section*{Acknowledgment}
This work was partially  supported by the French National Agency for Research (ANR -08-EMER-009-01).
{\small 
\bibliographystyle{plain}

\begin{thebibliography}{10}

\bibitem{gauss:aghanim99}
N.~{Aghanim} and O.~{Forni}.
\newblock {Searching for the non-Gaussian signature of the CMB secondary
  anisotropies}.
\newblock {\em Astronomy and Astrophysics}, 347:409--418, July 1999.

\bibitem{gauss:aghanim03}
N.~{Aghanim}, M.~{Kunz}, P.~G. {Castro}, and O.~{Forni}.
\newblock {Non-Gaussianity: Comparing wavelet and Fourier based methods}.
\newblock {\em Astronomy and Astrophysics}, 406:797--816, August 2003.

\bibitem{rest:anscombe48}
F.J. Anscombe.
\newblock The transformation of {P}oisson, binomial and negative-binomial data.
\newblock {\em Biometrika}, 15:246--254, 1948.

\bibitem{bernardeau2002}
F.~{Bernardeau} and J.~{Uzan}.
\newblock {Non-Gaussianity in multifield inflation}.
\newblock {\em Physical Review D}, 66:103506--+, November 2002.

\bibitem{starck:bobin_herschel}
J.~{Bobin}, J.-L. {Starck}, and R.~{Ottensamer}.
\newblock {Compressed Sensing in Astronomy}.
\newblock {\em ArXiv e-prints}, 802, February 2008.

\bibitem{bouchet88}
F.~R. {Bouchet}, D.~P. {Bennett}, and A.~{Stebbins}.
\newblock Patterns of the cosmic microwave background from evolving string
  networks.
\newblock {\em Nature}, 335:410, 1988.

\bibitem{cur:demanet06}
E.~Cand\`es, L.~Demanet, D.~Donoho, and L.~Ying.
\newblock Fast discrete curvelet transforms.
\newblock {\em SIAM Multiscale Model. Simul.}, 5/3:861--899, 2006.

\bibitem{CandesDonohoCurvelets}
E.~J. Cand\`es and D.~L. Donoho.
\newblock Curvelets -- a surprisingly effective nonadaptive representation for
  objects with edges.
\newblock In A.~Cohen, C.~Rabut, and L.L. Schumaker, editors, {\em Curve and
  Surface Fitting: Saint-Malo 1999}, Nashville, TN, 1999. Vanderbilt University
  Press.

\bibitem{cur:candes99_1}
E.J. Cand\`es and D.~Donoho.
\newblock Ridgelets: the key to high dimensional intermittency?
\newblock {\em Philosophical Transactions of the Royal Society of London A},
  357:2495--2509, 1999.

\bibitem{CRT:cs}
Emmanuel Cand\`es, Justin Romberg, , and Terence Tao.
\newblock Robust uncertainty principles: Exact signal reconstruction from
  highly incomplete frequency information.
\newblock {\em IEEE Trans. on Information Theory}, 52(2):489--509, 2006.

\bibitem{Coifman01noiselets}
R.~Coifman, F.~Geshwind, and Y.~Meyer.
\newblock Noiselets.
\newblock {\em Appl. Comput. Harmon. Anal.}, 10(1):27--44, 2001.

\bibitem{CombettesWajs05}
P.~L. Combettes and V.~R. Wajs.
\newblock Signal recovery by proximal forward-backward splitting.
\newblock {\em SIAM Journal on Multiscale Modeling and Simulation},
  4(4):1168--1200, 2005.

\bibitem{donoho:cs}
D.~Donoho.
\newblock Compressed sensing.
\newblock {\em IEEE Trans. on Information Theory}, 52(4):1289--1306, 2006.

\bibitem{cur:donoho_01b}
D.L. Donoho and X.~Huo.
\newblock Uncertainty principles and ideal atomic decomposition.
\newblock {\em IEEE Transactions on Information Theory}, 47:2845--2862, 2001.

\bibitem{inpainting:elad05}
M.~{Elad}, J.-L. {Starck}, P.~{Querre}, and D.L. {Donoho}.
\newblock {Simultaneous Cartoon and Texture Image Inpainting using
  Morphological Component Analysis (MCA)}.
\newblock {\em J. on Applied and Computational Harmonic Analysis},
  19(3):340--358, 2005.

\bibitem{starck:jalal06}
M.J. Fadili, J.-L Starck, and F.~Murtagh.
\newblock Inpainting and zooming using sparse representations.
\newblock {\em The Computer Journal}, 2006.
\newblock submitted.

\bibitem{starck:jin05}
J.~Jin, J.-L. Starck, D.L. Donoho, N.~Aghanim, and O.~Forni.
\newblock Cosmological non-gaussian signatures detection: Comparison of
  statistical tests.
\newblock {\em Eurasip Journal}, 15:2470--2485, 2005.

\bibitem{starck:mur95_2}
F.~Murtagh, J.-L. Starck, and A.~Bijaoui.
\newblock Image restoration with noise suppression using a multiresolution
  support.
\newblock {\em Astronomy and Astrophysics, Supplement Series}, 112:179--189,
  1995.

\bibitem{pires09}
S.~{Pires}, J.~. {Starck}, A.~{Amara}, R.~{Teyssier}, A.~{Refregier}, and
  J.~{Fadili}.
\newblock {FASTLens (FAst STatistics for weak Lensing) : Fast method for Weak
  Lensing Statistics and map making}.
\newblock {\em AA}, 2009.
\newblock in press.

\bibitem{riazuelo2002}
A.~{Riazuelo}, J.-P. {Uzan}, R.~{Lehoucq}, and J.~{Weeks}.
\newblock Simulating cosmic microwave background maps in multi-connected
  spaces.
\newblock astro-ph/0212223, 2002.

\bibitem{starck:sta03_1}
J.-L. Starck, N.~Aghanim, and O.~Forni.
\newblock Detecting cosmological non-gaussian signatures by multi-scale
  methods.
\newblock {\em Astronomy and Astrophysics}, 416:9--17, 2004.

\bibitem{starck:sta95_1}
J.-L. Starck, A.~Bijaoui, and F.~Murtagh.
\newblock Multiresolution support applied to image filtering and deconvolution.
\newblock {\em CVGIP: Graphical Models and Image Processing}, 57:420--431,
  1995.

\bibitem{starck:sta01_3}
J.-L. Starck, E.~Cand\`es, and D.L. Donoho.
\newblock The curvelet transform for image denoising.
\newblock {\em IEEE Transactions on Image Processing}, 11(6):131--141, 2002.

\bibitem{starck:sta02_3}
J.-L. Starck, E.~Candes, and D.L. Donoho.
\newblock Astronomical image representation by the curvelet tansform.
\newblock {\em AA}, 398:785--800, 2003.

\bibitem{Starck2006}
J.-L. Starck, M.J. Fadili, and F.~Murtagh.
\newblock {T}he {U}ndecimated {W}avelet {D}ecomposition and its
  {R}econstruction.
\newblock {\em IEEE Transactions on Image Processing}, 16(2):297--309, 2007.

\bibitem{starck:sta05_2}
J.-L. Starck, Y.~Moudden, P.~Abrial, and M.~Nguyen.
\newblock Wavelets, ridgelets and curvelets on the sphere.
\newblock {\em Astronomy and Astrophysics}, 446:1191--1204, 2006.

\bibitem{starck:sta98_3}
J.-L. Starck and F.~Murtagh.
\newblock Automatic noise estimation from the multiresolution support.
\newblock {\em Publications of the Astronomical Society of the Pacific},
  110:193--199, 1998.

\bibitem{starck:book02}
J.-L. Starck and F.~Murtagh.
\newblock {\em Astronomical Image and Data Analysis}.
\newblock Springer-Verlag, 2002.

\bibitem{starck:book06}
J.-L. {Starck} and F.~{Murtagh}.
\newblock {\em {Astronomical Image and Data Analysis}}.
\newblock Astronomical image and data analysis, by J.-L.~Starck and
  F.~Murtagh.~Astronomy and astrophysics library.~ Berlin: Springer, 2006,
  2006.

\bibitem{starck:book98}
J.-L. Starck, F.~Murtagh, and A.~Bijaoui.
\newblock {\em Image Processing and Data Analysis: The Multiscale Approach}.
\newblock Cambridge University Press, 1998.

\bibitem{sunyaev80}
R.~A. {Sunyaev} and I.~B. {Zeldovich}.
\newblock {Microwave background radiation as a probe of the contemporary
  structure and history of the universe}.
\newblock {\em Annual Review of Astronomy and Astrophysics}, 18:537--560, 1980.

\bibitem{starck:zhang07}
B.~Zhang, M.J. Fadili, and J.-L. Starck.
\newblock Wavelets, ridgelets and curvelets for poisson noise removal.
\newblock {\em IEEE Transactions on Image Processing}, 17(7):1093--1108, 2008.

\end{thebibliography}

\begin{IEEEbiography}[{
=1.25in,clip,keepaspectratio]{starck.eps}}]{Jean-Luc Starck} 
Jean-Luc Starck has a Ph.D from University
Nice-Sophia Antipolis and an Habilitation from University
Paris XI. He was a visitor at the European Southern
Observatory (ESO) in 1993, at UCLA in 2004 and at Stanford's statistics
department in 2000 and 2005. He has been a Researcher at CEA
since 1994. His research interests include image processing,
statistical methods in astrophysics and cosmology.
He is an expert in multiscale methods such wavelets and curvelets,
He is leader of the project Multiresolution at CEA and he is
a core team member of the PLANCK ESA project.
He has published more than 100 papers in different areas in
scientific journals. He is also author of two books entitled
Image Processing and Data Analysis: the Multiscale Approach
(Cambridge University Press, 1998), and Astronomical Image and
Data Analysis (Springer, 2nd edition, 2006).
\end{IEEEbiography}

 \begin{IEEEbiography}[{
=1.25in,clip,keepaspectratio]{jbobin.eps}}]{Jerome 
Bobin} 
Jérôme Bobin graduated from the Ecole Normale Superieure (ENS) de Cachan, France, in 2005 and received the M.Sc. degree in signal and image processing from ENS Cachan and Université Paris XI, Orsay, France. He received the Agrégation de Physique in 2004. Since 2005, he is pursuing his Ph.D with J-L.Starck at the CEA. His research interests include statistics, information theory, multiscale methods and sparse representations in signal and image processing.
\end{IEEEbiography}
\end{document}